\documentclass[11pt,twoside]{article}
\usepackage{amsfonts}
\DeclareMathAlphabet{\mathsfsl}{OT1}{cmss}{m}{sl}
\DeclareMathAlphabet{\mathscr}{U}{eus}{m}{n}
\DeclareMathAlphabet{\matheur}{U}{eur}{m}{n}
\SetMathAlphabet{\matheur}{bold}{U}{eur}{b}{n}
\newcommand{\nn}{\nonumber \\}
\newcommand{\beq}{\begin{equation}} 
\newcommand{\eeq}{\end{equation}}
\newcommand{\bea}{\begin{eqnarray}}
\newcommand{\eea}{\end{eqnarray}}

\newcommand{\pa}{\partial}
\newcommand{\bruch}[2]{{\textstyle \frac{#1}{#2}}}
\newcommand{\A}{\alpha}
\newcommand{\B}{\beta}
\newcommand{\C}{\gamma}
\newcommand{\GA}{\Gamma}
\newcommand{\D}{\delta}
\newcommand{\dxy}{\delta ({\bf x},{\bf y})}

\newcommand{\E}{\varepsilon}

\newcommand{\M}{\mu}
\newcommand{\N}{\nu}

\newcommand{\OM}{\omega}
\newcommand{\OO}{\Omega}

\newcommand{\DC}{\mathcal{D}}

\newcommand{\JC}{\mathcal{J}}

\newcommand{\VC}{\mathcal{V}}

\newcommand{\oep}{\bar{\varepsilon}}

\newcommand{\PtA}{{\tilde P}^A}
\newcommand{\PtB}{{\tilde P}^B}
\newcommand{\bx}{{\bf x}}
\newcommand{\by}{{\bf y}}

\newcommand{\so}{{\rm SO}(1,2) \times {\rm SO}(16)}
\newcommand{\sx}{{\rm SO}(1,2) \times {\rm SO}(8)}
\newcommand{\su}{{\rm SO}(1,3) \times {\rm SU}(8)}

\begin{document}
\def\draft{\pagestyle{draft}\thispagestyle{draft}
\global\def\draftcontrol{1}}
\global\def\draftcontrol{0}
\arraycolsep3pt

\thispagestyle{empty}
\begin{flushright} hep-th/9709227
                   \\  AEI-045
\end{flushright}
\vspace*{1.0cm}
\begin{center}
 {\LARGE \sc New Canonical Variables\\[1ex]
             for $d=11$ Supergravity%
  }\\
 \vspace*{1cm}
 {\sl
     Stephan Melosch\footnotemark[1] and
     Hermann Nicolai\footnotemark[2] \\
 \vspace*{6mm}
     \footnotemark[1]
     II. Institut f\"ur Theoretische Physik, Universit\"at Hamburg\\
     Luruper Chaussee 149, D-22761 Hamburg, Germany\\
 \vspace*{3mm}
     \footnotemark[2]
     Max-Planck-Institut f\"ur Gravitationsphysik\\
     Albert-Einstein-Institut \\
     Schlaatzweg 1, D-14473 Potsdam, Germany} \\
 \vspace*{1cm}
\begin{minipage}{11cm}\footnotesize
\textbf{Abstract:}
A set of new canonical variables for $d=11$ supergravity is 
proposed which renders the supersymmetry variations and 
the supersymmetry constraint polynomial.
The construction is based on the $\so$ invariant 
reformulation of $d=11$ supergravity given in \cite{nic1}, 
and has some similarities with Ashtekar's reformulation of 
Einstein's theory. The new bosonic variables fuse the gravitational 
degrees of freedom with those of the three-index photon $A_{MNP}$ 
in accordance with the hidden symmetries of the dimensionally 
reduced theory. Although $E_8$ is not a symmetry of the theory, 
the bosonic sector exhibits a remarkable $E_8$ structure, hinting 
at the existence of a novel type of ``exceptional geometry''. 
\end{minipage}
\end{center}
\setcounter{footnote}{0}

Recent advances in string theory (see e.g. \cite{witten}) 
have lent renewed support to the long held belief that 
$d=11$ supergravity \cite{CJS} has a fundamental role to 
play in the unification of fundamental interactions. In this 
letter, we present an unconventional formulation of this 
theory, developing further the results of refs. \cite{dewnic1, nic1} 
where new versions of $d=11$ supergravity with local $\su$ and $\so$
tangent space symmetries, respectively, were presented. 
In both versions the supersymmetry variations were shown to 
acquire a polynomial form from which the corresponding formulas 
for the maximal supergravities in four and three dimensions can 
be read off directly and without the need for complicated duality 
redefinitions. Our reformulation can thus be regarded as a step 
towards the complete fusion of the bosonic degrees of freedom of $d=11$ 
supergravity (i.e. the elfbein and the antisymmetric tensor $A_{MNP}$)
in a way which is in harmony with the hidden symmetries 
of the dimensionally reduced theories \cite{CJ,julia}\footnote{In a
different context, the fusion of gravitational and matter
(Yang Mills) degrees of freedom was also attempted in 
\cite{Peldan}.}. The results are very suggestive of a novel
kind of ``exceptional geometry'' for $d=11$ supergravity (or some 
bigger theory containing it) that would be intimately tied 
to the special properties of the exceptional groups, and would be
characterized by relations such as (\ref{id1})--(\ref{id4}) below,
which have no analog in ordinary Riemannian geometry.

The hamiltonian formulation of our results reveals surprising
similarities with Ashtekar's reformulation of Einstein's
theory \cite{asht} (for a conventional hamiltonian treatment 
of $d=11$ supergravity, cf. \cite{Diaz}\footnote{The 
Chern-Simons part of the $d=11$ action was recently considered 
in \cite{smolin}.}). More specifically, the ``248-bein'' to be
introduced below is the analog of the inverse densitized
dreibein (or ``triad'') in \cite{asht}. Furthermore, in terms 
of the canonical variables proposed here the supersymmetry 
constraints become polynomial; the polynomiality 
of the remaining canonical constraints is then implied by
supersymmetry and the polynomiality of the canonical brackets.
Unfortunately, not all sectors of the theory are as simple 
as one might have wished, and a further simplification will very 
likely require a better understanding of the exceptional structures 
alluded to above, as well as the further extension of the results 
of \cite{dewnic1,nic1} to incorporate the even larger (infinite 
dimensional) symmetries arising in the dimensional reductions of 
$d=11$ supergravity to two and one dimensions, respectively.

Let us first recall the main results, conventions and 
notation of \cite{nic1} (further details will be provided 
in a forthcoming thesis \cite{Melosch}). To derive the new 
version from the original formulation of $d=11$ supergravity, 
one first breaks the original tangent space symmetry SO(1,10) to 
its subgroup $\sx$ through a partial choice of 
gauge for the elfbein, and subsequently enlarges it again to 
$\so$ by introducing new gauge degrees of freedom. 
This symmetry enhancement requires suitable redefinitions 
of the bosonic and fermionic fields, or, more succinctly, 
their combination into tensors w.r.t. the new tangent space 
symmetry. The basic strategy underlying this construction goes back to
\cite{CJ}, but the crucial difference is that the dependence 
on {\it all eleven} coordinates is retained here.
The construction thus requires a 3+8 split of the $d=11$ coordinates 
and indices, implying a similar split for all tensors of the theory. 
Accordingly, curved $d=11$ indices are decomposed as 
$M=(\mu ,m), N=(\nu,n),...$ with $\M,\N,\ldots=0,1,2$ and 
$m,n,\ldots = 3,...,10$; the associated flat indices are denoted 
by $\A,\B,\ldots$ and $a,b,\ldots$, respectively. To distinguish the  
flat index 0 from its curved homolog, we will use the label 
$t$ for the latter. We will occasionally refer to the $d=8$ 
coordinates and field components as ``internal'' coordinates and 
components, respectively. We will also need SO(16) indices $I,J,...$ 
for the 16-dimensional vector representation, and dotted 
and undotted indices $\dot{A}, \dot{B},...$ and $A,B,...$ for the 
128-dimensional left and right handed spinor representations, 
respectively; antisymmetrized index pairs $[IJ]$ label the 
120-dimensional adjoint representation.

In the fermionic sector, the symmetry breaking induces a 
split of the 32 components of an SO(1,10) Majorana spinor in
accordance with the subgroup SO(1,2)$\, \times \,$SO(8); these fermions 
are then reassembled into $2\times 16$ component Majorana spinors  
transforming under the new local $\so$ symmetry.
For notational simplicity, we will suppress the SO(1,2) spinor
indices throughout. In this manner, the $d=11$ gravitino
--- the only fermionic field in $d=11$ supergravity --- is decomposed 
into a ``gravitino-like'' vector spinor $\Psi_{\M}^{I}$ 
belonging to the $\mathbf{16}$ vector representation of SO(16) 
and a ``matter-like'' fermion field $\chi_{\dot{A}}$ which is assigned 
to the (left-handed) $\mathbf{128}_c$ spinor representation;
these are also the fermionic representations appearing in the 
dimensionally reduced theory, i.e. $N=16$ supergravity in 
three dimensions \cite{marcus}. 

In the bosonic sector, the elfbein and the three-index photon
are combined into new objects covariant w.r.t. to $d=3$ coordinate
reparametrizations and the new $\so$ tangent space 
symmetry. The elfbein contains the (Weyl rescaled) dreibein
$e_{\M}{}^{\A}$ and the Kaluza Klein vector $B_\mu{}^{m}$; the
remaining ``matter-like'' degrees of freedom are merged into a 
rectangular ``248-bein'' $(e^{m}_{IJ},e^{m}_{A})$ obeying a 
generalization of the usual vielbein postulate. This 248-bein, which 
in the reduction to three dimensions contains all the propagating 
bosonic matter degrees of freedom of $d=3,N=16$ supergravity,
is very much analogous to part of Ashtekar's variables: in a 
special SO(16) gauge, it is identified with the inverse densitized
internal achtbein contracted with a $\Gamma$-matrix. Because
$([IJ],A)$ label the 248-dimensional adjoint representation
of $E_8$, it is really a soldering form that relates upper (internal) 
world indices to $E_8$ indices. In addition, we have the composite
fields $(Q_{\M}^{IJ}, P_{\M}^{A})$ and $(Q_{m}^{IJ}, P_{m}^{A})$,
which make up an $E_8$ connection in {\it eleven} dimensions and
whose explicit expressions in terms of the $d=11$ coefficients of
anholonomity and the four-index field strength $F_{MNPQ}$ can be 
found in \cite{nic1}. 

There are various algebraic constraints between the vielbein
components which ensure that the total number of bosonic physical 
degrees of freedom is the same as before, and which are most
conveniently proven in the special SO(16) gauge of \cite{nic1}.
These relations are without analog in ordinary Riemannian 
geometry, because they rely in an essential way on special properties 
of the exceptional group $E_8$. We have
\bea
e^m_A e^n_A - \bruch{1}{2}e^m_{IJ} e^n_{IJ} = 0 \label{id1}
\eea
and
\bea
\GA^{IJ}_{AB} \Big( e^m_B e^n_{IJ} - e^n_B e^m_{IJ} \Big) &=& 0  \nn
\GA^{IJ}_{AB} e^m_A e^n_B + 4 e^m_{K[I} e^n_{J]K}  &=&  0 \label{id2}
\eea
where, of course, $\Gamma^I_{A\dot A}$ are the standard SO(16) 
$\GA$-matrices and $\GA_{AB}^{IJ}\equiv (\GA^{[I} \GA^{J]})_{AB}$, etc.
The identities (\ref{id2}) were already given in \cite{nic4}; the minus 
sign in (\ref{id1}) reflects the fact that we are dealing with
the non-compact form $E_{8(+8)}$. In addition we will need 
to make use of the more general relations
\bea
2 e^{(m}_{A} e^{n)}_{B} 
&=& - \bruch{1}{16} (\GA^{IJ}\GA^{KL})_{AB} e_{IJ}^{m} e_{KL}^{n}
  + \bruch{1}{8} \GA^{IJ}_{AC}e_{C}^{m} \ e_{D}^{n}\GA^{IJ}_{DB}  \nn
 2 e^{(m}_{IJ} e^{n)}_{KL} &=&
     - 4 e^{(m}_{I[K} e^{n)}_{L]J}  
     + 2 e^{m}_{IM} e^{n}_{M[K} \D_{L]J}  
      - 2 e^{n}_{JM} e^{m}_{M[K} \D_{L]I} \nn
 && - \bruch{1}{4} e^{m}_{A} (\GA^{IJ}\GA^{KL})_{AB} e^{n}_{B}  \nn
2 e^{(m}_{A} e^{n)}_{IJ}  &=&
 - \GA_{AB}^{IK} e^{(m}_{B} e^{n)}_{KJ}
 + \GA_{AB}^{JK} e^{(m}_{B} e^{n)}_{KI}  
  - \bruch{1}{8} (\GA^{IJ}\GA^{KL})_{AB} e^{m}_{B} e^{n}_{KL} 
\label{id3}
\eea
where $(...)$ denotes symmetrization with strength one (the 
combinations antisymmetric in $[mn]$ contained in the above 
expressions all reduce to (\ref{id2}) and thus yield no new
information). While the SO(16) covariance of these 
equations is manifest, it turns out, remarkably, that they are also 
covariant (i.e. transform into one another) under $E_8$. Obviously, 
(\ref{id1}) and (\ref{id2}) correspond to the singlet and the adjoint 
representations of $E_8$. The relations (\ref{id3}) are not 
irreducible as they stand, but can be made so by projecting
out the singlet and the $\bf 3785$ representation of $E_8$;
the latter is given by 
\bea
e^{(m}_{IK} e^{n)}_{JK} - \bruch{1}{16} \delta_{IJ} 
e^m_{KL} e^n_{KL}  &=& 0   \nn
\GA^K_{\dot A B} e^{(m}_B e^{n)}_{IK} - \bruch{1}{14}
\GA^{IKL}_{\dot A B} e^{(m}_B e^{n)}_{KL} &=& 0 \nn
e^{(m}_{[IJ} e^{n)}_{KL]} + \bruch{1}{24}
e^m_A \GA^{IJKL}_{AB} e^n_B  &=& 0 \label{id4}
\eea
After this projection, one is left with the $\bf 27000$ representation 
of $E_8$. There are presumably more algebraic relations of 
this type, but we have made no attempts towards a complete 
classification\footnote{The 56-bein 
$(e^m_{AB},e^{mAB})$ of \cite{dewnic1} obeys analogous relations, 
which can be similarly assigned to irreducible representations 
of $E_7$. For instance, the analog of (\ref{id1}) reads 
$$ 
e^m_{AB} e^{nAB} - e^{mAB} e^n_{AB} = 0
$$
where the combination on the l.h.s. is just the symplectic second order 
invariant for the 56 dimensional fundamental representation of $E_7$. 
Furthermore, we have 
$$
e^m_{AC} e^{nBC} + e^n_{AC} e^{mBC} - 
   \bruch{1}{4} \delta_A^B e^m_{CD} e^{nCD} = 0
$$
$$
e^m_{[AB} e^n_{CD]} - \bruch{1}{24} \varepsilon_{ABCDEFGH}
  e^{mEF} e^{nGH} = 0
$$
corresponding to the $\bf 133$ representation of $E_7$. The first 
of these relations was referred to as the ``Clifford property''
in \cite{dewnic1} (cf. eq. (3.9)).}. 

The 248-bein and the new connection fields are subject to 
the so-called ``generalized vielbein postulates" 
\bea
 D_{\mu} e_{IJ}^{m} +
\partial_{n} B_{\mu}{}^{n} e^{m}_{IJ} +
\partial_{n}B_{\mu}{}^{m} e^{n}_{IJ} +
P_{\mu}^{A}\Gamma^{IJ}_{AB} e^m_B & = & 0  \nn
D_{\mu} e_{A}^{m} +
\partial_{n} B_{\mu}{}^{m} e^{n}_{A} +
\partial_{n}B_{\mu}{}^{n} e^{m}_{A}
-\bruch{1}{2} \Gamma^{IJ}_{AB} P_{\mu}^{B} e^{m}_{IJ} & = & 0
\label{VVP3}
\\
D_{m} e^{n}_{IJ}
+P_{m}^{A} \Gamma^{IJ}_{AB} e^n_B & = & 0 \nn
D_{m} e^{n}_{A}
-\bruch{1}{2} \Gamma^{IJ}_{AB} P_{m}^{B} e^{n}_{IJ} & = & 0
\label{VVP4}
\eea
where the SO(16) covariant derivatives on the 248-bein are defined by
\bea
D_\mu e^n_{IJ} &=&
\DC_{\M} e^n_{IJ} + 
   2{{Q_\mu}^K}_{[I} e^{n}_{J]K} \nn
D_\mu e^n_A  &=&
\DC_{\M} e^n_{A} 
+ \bruch{1}{4} Q_{\M}^{IJ} \GA_{AB}^{IJ} e^{n}_{B} \\
D_m e^n_{IJ}  &=& 
\pa_m e^n_{IJ} + 2{{Q_m}^K}_{[I}e^{n}_{J]K}  \nn
D_m e^n_A    &=&   
\pa_m e^n_A + \bruch{1}{4} Q_{m}^{IJ} \GA_{AB}^{IJ} e^{n}_{B}
\eea
with the modified derivative 
\beq
\DC_{\M} := \pa_{\M} - B_\mu{}^{m} \pa_{m}
\eeq
Like (\ref{id1})--(\ref{id4}), these relations are $E_8$ covariant;
in fact, (\ref{VVP3}) and (\ref{VVP4}) simply state the covariant 
constancy of the 248-bein w.r.t. to a fully $E_8$ covariant 
derivative. This feature is reminiscent of the covariant constancy 
of the inverse densitized dreibein in $d=4$ (canonical) gravity w.r.t. 
a covariant derivative involving only the Ashtekar connection.

Unlike the previous relations, the following duality constraint
has no analog in the $\su$ version of \cite{dewnic1}: 
\beq
e^m_A P_\mu^A = {\varepsilon_\mu}^{\nu\rho}
\DC_\nu B_\rho{}^m   \label{dual1}
\eeq
In the reduction to $d=3$, the relation (\ref{dual1})
shows how the Kaluza Klein vectors $B_\mu{}^m$ are dualized 
into scalar fields, but a straightforward dualization is obviously
no longer possible if the dependence on all eleven coordinates 
is retained, due to the explicit appearance of the field $B_\mu{}^m$
without a derivative in this and other formulas. 

The invariance of $d=11$ supergravity under local supersymmetry
in its original form implies an analogous local supersymmetry
for the new formulation as well. Deriving the new transformation
laws requires some tedious calculations, which are most conveniently
done ``backwards'' as explained in \cite{dewnic1,nic1}. In particular, 
attention must be paid to various compensating rotations, and the 
final variations differ from those of the original theory by a local
SO(16) rotation. Modulo this compensating rotation, the supersymmetry 
variations of the bosonic fields assume a rather simple form \cite{nic1}, 
viz.\footnote{We have changed a few normalizations
in comparison with \cite{nic1}.}
\bea
\D e_{\mu}{}^{\A} & = &
\bruch {1}{2} \oep^{I} \gamma^{\A} \Psi_{\mu}^{I}  \label{vare} \nn
\D B_{\mu}{}^{m}  & = &
\bruch{1}{2} e^{m}_{IJ} \oep^{I} \Psi_{\mu}^{J}
+ \bruch{1}{4} e_{A}^{m} \Gamma^{I}_{A\dot{A}} \oep^{I} \gamma_{\mu}
\chi_{\dot{A}}  \label{varB} \nn
\D e^{m}_{IJ} & = & e^m_A \GA^{IJ}_{AB} \omega^B   \quad , \quad
\delta e^{m}_{A} = \bruch{1}{2} e^m_{IJ} \GA^{IJ}_{AB} \omega^B 
               \label{var248}
\eea
where the last two variations have been written in the form 
of a local $E_8/$SO(16) rotation with parameter 
\bea
\omega^A:= \bruch{1}{4}\oep^I \GA^I_{A\dot A}\chi_{\dot A} \label{om}
\eea
The fact that the variations on the 248-bein take this
form ensures the compatibility of the algebraic constraints
(\ref{id1})--(\ref{id4}) with local supersymmetry. 
The fermionic fields transform as
\bea
\delta \Psi_{\mu}^{I} & = &
(D_{\mu} - \bruch{1}{2} \partial_{m} B_{\mu}{}^{m})\varepsilon^{I} +
   \nn   && +
\bruch{1}{2} e^{m}_{IJ} \Big( \gamma_{\mu} D_{m} \varepsilon^{J}
+  D_{m}(\gamma_{\mu} \varepsilon^{J})\Big)
    + \bruch{1}{2} \gamma_{\mu}
      e_{A}^{m} \Gamma^{IJ}_{AB} P_{m}^{B} \varepsilon^{J} \label{varpsi} \nn
\delta \chi_{\dot{A}} & = & 
4 \Gamma^{I}_{\dot{A}A} \gamma^{\mu} \varepsilon^{I} P_{\mu}^{A} +
  \nn  &&  + \,
2 \Gamma^{I}_{\dot{A}A} e^{m}_{A} D_{m} \varepsilon^{I}
+ \bruch{1}{2} e^{m}_{IJ} \left( \Gamma^{IJ} \Gamma^K
   + 4 \Gamma^I \delta^{JK} \right)_{\dot{A} A}
                  P_{m}^{A} \varepsilon^{K}   \label{varchi}
\eea
modulo higher order fermionic contributions. The $\so$
covariant derivatives on the spinors are given by
\bea
D_\mu \varepsilon^I & = & 
(\DC_{\M} + \bruch{1}{2} 
 A_\mu{}^{\A} \C_{\A}) \varepsilon^I + Q_{\M}^{IJ} \varepsilon^I
 \\
D_m \varepsilon^I & = & (\pa_{m} + \bruch{1}{2} A_m{}^{\A} \C_{\A})
\varepsilon^I +   Q_{m}^{IJ} \varepsilon^I
\\
D_\mu \chi_{\dot{A}} & = & 
(\DC_{\M} + \bruch{1}{2}
 A_\mu{}^{\A} \C_{\A}
) \chi_{\dot{A}} 
+ \bruch{1}{4}  Q_{\M}^{IJ} \GA_{\dot{A}\dot{B}}^{IJ} \chi_{\dot{B}}
\\
D_m \chi_{\dot{A}} & = & 
(\pa_{m}  + \bruch{1}{2} A_m{}^{\A} \C_{\A})\chi_{\dot{A}} 
+ \bruch{1}{4}  Q_{m}^{IJ} \GA_{\dot{A}\dot{B}}^{IJ} \chi_{\dot{B}}
\eea
where we have introduced the dualized connections 
\beq
A_m{}^{\A} := - \bruch{1}{2} \E^{\A\B\C} e_{\B}^{~\N}\pa_{m} e_{\N\C}
\eeq
 and 
\beq
A_\mu{}^{\A} :=  \bruch{1}{2} \E^{\A\B\C} \Big({\OM}_{\mu \, \B\C} +
2  e_{\mu \B} e_{\C}{}^\nu \pa_m B_\nu{}^m \Big)
\eeq
The SO(1,2) spin connections 
\beq
{\OM}_{\mu \, \A\B} = \bruch{1}{2} e_{\M}{}^{\C} 
\left( \OO_{\A\B\C} - \OO_{\B\C\A} - \OO_{\C\A\B} \right)
\eeq
and the anholonomity coefficients 
\beq
\OO_{\A\B\C} = 2 e_{[\A}{}^{\M}  e_{\B]}{}^{\N} \DC_{\M} e_{\N\C}
\eeq
differ from the standard $d=3$ expressions by their extra dependence 
on $B_\mu{}^m$ and the internal coordinates. In addition to 
their manifest covariance properties the above variations are 
evidently polynomial in the new fields.

We have also calculated the variations of the connection fields,
which were not given in \cite{dewnic1,nic1},
\bea 
\D P_{\M}^{A} & = & D_\mu \omega^A
- \D B_\mu{}^m P_m^{A}
+ e^{m}_{A} X_{m\M}  \nn & &  
-\bruch{1}{32} e^{m}_{IJ}  (\GA^{IJ}\GA^{K})_{A \dot{B}}
  D_{m}(\oep^{K}\gamma_{\M}\chi_{\dot{B}})   \nn &&
+ \bruch{1}{32} \GA^{IJ}_{AB} e^m_{B} P_m^{C} 
   (\GA^{IJ}\GA^{K})_{C\dot{D}}\oep^{K}\gamma_{\M}\chi_{\dot{D}} 
\nn & & 
- \bruch{1}{8} \GA^{IJ}_{AB} e^{m}_{B} D_{m}(\oep^{I} \Psi_{\M}^{J}) 
- \bruch{1}{16} (\GA^{IJ}\GA^{KL})_{AB} P_m^{B} e^m_{IJ} 
\oep^{K} \Psi_{\M}^{L}
\label{varPmu} \eea
\bea
\D Q_{\M}^{IJ} & = &
      2 \, P_\mu^{A} \GA^{IJ}_{AB} \omega^B 
 - \D B_\mu{}^m Q_m^{IJ}
+ 2 e^{m}_{IJ} X_{m\M}
\nn & &
- \bruch{1}{8}  e^{m}_{A} (\GA^{IJ}\GA^{K})_{A \dot{B}}
 D_{m}( \oep^{K}\gamma_{\M} \chi_{\dot{B}}) 
\nn & &
- \bruch{1}{4} e^m_{IK}P_m^{A}(\GA^{KJ}\GA^{L})_{A\dot{B}}
  \oep^{L}\gamma_{\M}\chi_{\dot{B}}
+ \bruch{1}{4} e^m_{JK}P_m^{A}(\GA^{KI}\GA^{L})_{A\dot{B}}
  \oep^{L}\gamma_{\M}\chi_{\dot{B}}
\nn & &
   - e^{m}_{IK} D_{m}(\oep^{[J} \Psi_{\M}^{K]}) 
   + e^{m}_{JK} D_{m}(\oep^{[I} \Psi_{\M}^{K]}) 
\nn & &
- \bruch{1}{4} e_{A}^{m} (\GA^{IJ}\GA^{KL})_{AB} P_m^{B} 
\oep^{K} \Psi_{\M}^{L} 
\label{varQmu} \eea
with the parameter (\ref{om}). These transformations have been 
computed by varying (\ref{VVP3}) and making use of the identities 
(\ref{id3}) to solve for $\delta P_\mu^A$ and 
$\delta Q_\mu^{IJ}$. The contributions here which are not of 
the form of a local $E_8$ rotation originate without exception 
from the variation of $B_\mu{}^m$; in particular, the terms 
$\delta B_\mu{}^m Q_m^{IJ}$ and $\delta B_\mu{}^m P_m^A$
are needed to maintain covariance. The remaining ambiguity 
is contained in the terms proportional to $X_{m\mu}$; 
these contributions drop out in (\ref{VVP3}) and must 
be determined directly by comparison with the original 
variations of $d=11$ supergravity. A subtlety here, not encountered 
up to now, is that the variations agree with those of the original 
$d=11$ theory only upon use of the Rarita-Schwinger equation, 
i.e. on shell. This feature can be directly traced to the 
occurrence of dualized bosonic field strengths in the explicit 
expressions for $P_\mu^A$ and $Q_\mu^{IJ}$. 

By contrast, the supersymmetry variations of the ``internal'' 
components $\delta P_m^A$ and $\delta Q_m^{IJ}$ that follow from the 
generalized vielbein postulate (\ref{VVP4}) take a much simpler form
\bea
\delta P_m^A & = & D_m \omega^A + e^{n}_{A} X_{mn}
\nn
\delta Q_m^{IJ} & = & 2 \, P_m^A \GA^{IJ}_{AB} \omega^B 
+ 2 e^{n}_{IJ} X_{mn}  \label{varPm}
\eea
Again there is an ambiguity: the terms proportional to $X_{mn}$
drop out of the variation of (\ref{VVP4}) (to see this, use
must be made of (\ref{id2})). However, inspection shows that no 
terms with this index structure and the correct dimensionality 
can be manufactured out of the $\so$ covariant fields, and 
we therefore conclude that $X_{mn}$ must vanish. 

With the new SO(16) fields at hand it is possible to rewrite the
fermionic part of the $d=11$ Lagrangian in this environment \cite{Melosch}.
This is sufficient for the derivation of the supersymmetry constraint, 
which is just the time component of the Rarita Schwinger equation 
expressed in terms of the canonical variables. To determine the
full $\so$ invariant Lagrangian is much harder: because of the 
explicit occurrence of the three index field $A_{MNP}$ in the 
original Lagrangian of \cite{CJS}, one cannot directly rewrite 
the latter, but must go back to the $\so$ covariant bosonic 
equations of motion. Since we are here solely concerned with 
the supersymmetry constraint we refer readers to \cite{Melosch} 
for a detailed discussion. 

In setting up the hamiltonian formulation we follow the
standard procedure (see e.g. \cite{CG}), which requires
amongst other things that we foliate the $d=11$ space time 
by spatial slices. This entails in particular that the 
latter must be assumed to be of the global 
form ${\bf R} \times \Sigma_{10}$. The ten-dimensional
spatial manifold $\Sigma_{10}$ is locally parametrized by   
coordinates ${\bf x},{\bf y}$,..., so that the $d=11$ coordinates 
are represented as $x^M =(t,{\bf x})$, etc.; $d=3$ indices will be 
split as $\mu =(t,i)$ with $i,j,...= 1,2$. In order to determine 
the canonical variables ``from scratch'' we would again need 
to know the full $\so$ invariant Lagrangian. Since only its
fermionic part is available, we will proceed in a more 
pragmatical fashion, requiring that the bosonic brackets lead to 
the correct supersymmetry variations on all canonical fields.
The correctness of the ensuing brackets can be tested in 
alternative ways, for instance by inspection of the composite 
connections and their field content in terms of the original 
$d=11$ fields. Furthermore, in the dimensional reduction to $d=3$, 
the canonical brackets must match with those derived in \cite{nic4}.

The fermionic brackets are easily deduced from the fermionic 
Lagrangian. Because of the Majorana conditions, 
we have second class constraints such as 
\bea
\pi_{\dot A} = \bruch{1}{4} e e_\alpha{}^t \gamma^\alpha \chi_{\dot A} 
\eea
for the ``matter fermions'' (note that 
$e e_\alpha{}^t \equiv \varepsilon_{\alpha \beta \gamma}
\varepsilon^{ij} e_i{}^\beta e_j{}^\gamma$). These constraints 
are dealt with in the usual fashion, and after a little algebra 
we arrive at the following equal time (Dirac) brackets
\bea
\{\Psi_{i}^{I}({\bf x}),\bar{\Psi}_{j}^{J}({\bf y})\} &=&
\varepsilon_{ij} \D^{IJ} \dxy      \nonumber       \\
\{\chi_{\dot{A}} ({\bf x}) ,\bar{\pi}_{\dot{B}}({\bf y})\} &=& 
\D_{\dot{A}\dot{B}} \dxy
\eea

Owing to the complications described above, the canonical
structure of the bosonic sector is considerably more involved. 
First of all, because of the algebraic and differential constraints 
on the 248-bein and the connections given before, our new bosonic
variables constitute a redundant set, and thus cannot be grouped 
into canonically conjugate pairs. In principle, one could 
eliminate these redundancies by solving the constraints, but this
would force us to abandon the new local symmetries, and thereby 
obscure the new geometrical structures we are about to expose. 
For this reason, we will keep the constraints; their consistency 
with the canonical brackets essentially hinges on the $E_8$ 
structure of the bosonic sector.

Secondly, in the original version of $d=11$ supergravity, 
all fields are gauge connections associated with the 
local symmetries. Their time components thus serve as Lagrange 
multipliers for the associated constraint generators. Of these, 
the time components of the dreibein and the gravitino survive 
as Lagrange multipliers of the Hamiltonian and diffeomorphism
constraints, and the supersymmetry constraint, respectively. 
The time components $B_t{}^m$ correspond to local reparametrizations
of the ``internal'' coordinates, and their role is somewhat obscured
in the present framework. Since the original 
local SO(1,10) symmetry has been traded for $\so$, of the 
original $d=11$ spin connection, only the component $A_t{}^\alpha{}$ 
remains as a Lagrange multiplier for local SO(1,2). Instead of the 
``internal'' part of the spin connection, 
we now have the SO(16) constraint $\phi^{IJ}$ multiplying the time 
component $Q_t^{IJ}$ of the SO(16) gauge connection. The former 
generates local SO(16) rotations on all fields, viz.
\bea
\delta_\omega \varphi ({\bf x}) =
\Big\{ \varphi ({\bf x}) , \bruch{1}{2} \int d{\bf y} \, 
\omega^{IJ}({\bf y}) \phi^{IJ} ({\bf y}) \Big\}
\eea              
where $\omega^{IJ}$ is a local SO(16) parameter.

Thirdly, in terms of the original $d=11$ fields, the set of 
bosonic canonical variables consists of both ``elementary'' and 
``composite'' objects. Because of the dualizations implicit 
in our reformulation, the $\mu=i$ components of the 
$E_8$ connections contain time derivatives of the original 
fields. Consequently, there is not much point in singling out 
particular components as canonical momenta; however, we find it 
useful to define 
\beq
\tilde{P}^{A} := 4  e g^{t\M} P_{\M}{}^{A} 
\eeq
which in the reduction to three dimensions are just the canonical
momenta associated with the scalar fields, cf. \cite{nic4}.
We emphasize that bosonic brackets may vanish only up to bilinear 
fermionic contributions which we have neglected here; 
this may necessitate redefinitions by fermionic bilinears, 
such as e.g. for $A_i{}^\alpha$ in order to achieve 
$\{ A_i{}^\alpha , A_j{}^\beta \} = 0$ \cite{MN}.  

Apart from the brackets involving the connections, we have
\bea 
\{e_{i}{}^{\alpha}({\bf x}),A_{j}{}^{\beta}({\bf y})\} &=& 
\varepsilon_{ij} \eta^{\alpha\beta} \, \dxy  
\label{bos1} \\
\{B_i{}^{m}({\bf x}),P_{j}^{A}({\bf y}) \}
 & =&  \bruch{1}{4} \varepsilon_{ij} e^{m}_{A}({\bf x}) \, \dxy \nn
\{B_i{}^{m}({\bf x}),Q_{j}^{IJ}({\bf y}) \}
  &=&  \bruch{1}{2}  \varepsilon_{ij} e^{m}_{IJ}({\bf x})\, \dxy  
\label{bos3} \\
\{e^{m}_{A}({\bf x}),\tilde{P}^{B}({\bf y})\} &=&
   \bruch{1}{2}e^{m}_{IJ}({\bf x})\Gamma_{AB}^{IJ} \, \dxy   \nn
\{e^{m}_{IJ}({\bf x}),\tilde{P}^{B}({\bf y})\} &=&
               e^{m}_{A}({\bf x})\Gamma_{AB}^{IJ} \, \dxy
\label{bos5}
\eea
In the $d=3$ sector, the conjugate pairs $\varepsilon^{ik} e_k{}^\alpha$
and $A_j{}^\beta$ correspond to Ashtekar's variables 
in three dimensions \cite{Bengtsson} and become identical with 
them upon dimensional reduction to $d=3$. The second and third lines
extend this analogy to the Kaluza Klein components $B_i{}^m$. 
The non-vanishing result of the bracket between $B_i{}^m$ and
$(Q_j^{IJ}, P_j^A)$ is explained by the fact that latter fields
contain time derivatives of $B_i{}^m$ when expressed in terms 
of the original $d=11$ fields \cite{nic1}, as is also obvious
from (\ref{dual1}) by putting $\mu=i$. The consistency of the
algebraic identities (\ref{id1})--(\ref{id4}) with the above
brackets follows from their invariance under $E_8$ and the
fact that (\ref{bos5}) effectively corresponds to an 
$E_8$ rotation of the 248-bein.

The remaining brackets involving $\PtA$ are given by
\bea
\{ \PtA (\bx), \PtB (\by) \}  &=&   \GA_{AB}^{IJ} 
   \phi^{IJ} (\bx ) \dxy \nn
\{ \PtA (\bx), P_i^B(\by) \}  &=&  \Big(
  \delta^{AB}(\DC_i - \pa_m B_i{}^m(\bx)) +\bruch{1}{4} Q_i^{IJ}(\bx)
                 \GA^{IJ}_{AB}\Big) \dxy      \nn
\{ \PtA (\bx), Q_i^{IJ} (\by )\} &=& 
  2 \, \GA^{IJ}_{AB} P_i^B (\bx) \, \dxy    \nn
\{ \PtA (\bx), P_m^B(\by) \}  &=&  
         \Big(\D^{AB}\pa_m + \bruch{1}{4} Q_m^{IJ}(\bx)
                 \GA^{IJ}_{AB} \Big) \dxy      \nn
\{ \PtA (\bx), Q_m^{IJ} (\by )  \} &=& 
  2 \, \GA^{IJ}_{AB} P_m^B (\bx)\, \dxy    \label{brackets1}
\eea
where the derivative always acts on the first argument of the 
$\delta$-function. The brackets between the internal components
$P_m^A$ and $Q_m^{IJ}$ all vanish (this is not true for the
$d=3$ components). In the dimensional reduction to three dimensions, 
where $\pa_m\equiv 0$ and $Q_m^{IJ}=P_m^A\equiv 0$, these brackets 
coincide with the brackets predicted by the $\sigma$-model 
formulation \cite{nic4,MN} (modulo different normalizations). 

Varying the fermionic part of the action w.r.t. the time 
component of the gravitino and expressing the result in terms 
of the canonical variables yields the supersymmetry constraint
\bea
{\mathscr S}^{I} & = &
\varepsilon^{ij}\left( D_{i} - \bruch{1}{2}
         \partial_m B_i{}^m \right)\Psi_{j}^{I} 
 + \bruch{1}{4} \tilde{P}^{A} \Gamma^{I}_{A\dot{B}} \chi_{\dot{B}}
 +  \varepsilon^{ij}e_{j\alpha} \gamma^{\alpha} P_{i}^{A}
     \Gamma^{I}_{A\dot{B}} \chi_{\dot{B}}\nn
 &   & 
- \, 2 e^{m}_{A} \Gamma^{I}_{A\dot{B}} D_{m}\pi_{\dot{B}}
+ \bruch{1}{2} e^{m}_{JK}P_{m}^{A}
(\Gamma^{I}\Gamma^{JK} + 12 \delta^{IK}\Gamma^{J})_{A\dot{B}}
            \pi_{\dot{B}}  \nn
  &  & \nonumber
- \, \varepsilon^{ij} e_{j\alpha}
 \gamma^{\alpha} e_{IJ}^{m} D_{m} \Psi_{i}^{J}
-  \bruch{1}{2}\varepsilon^{ij}e_{j\alpha}
 \gamma^{\alpha} e_{A}^{m} \Gamma^{IJ}_{AB} P_{m}^{B} \Psi_{i}^{J} \nn
  && - \, \bruch{1}{2}\varepsilon^{ij}
 \gamma^{\alpha}  D_{m}e_{j\alpha} e^{m}_{IJ} \Psi^{J}_{i}
\label{Susy}  
\eea
The local supersymmetry of the theory is implemented by the constraint 
${\mathscr S}^I (\bx ) \approx 0$ in accordance with the general theory 
\cite{dirac}. As is well known, the supersymmetry constraint is
the key constraint because all other constraints can be obtained
from it by commutation. In this sense it is the ``square root'' of 
the bosonic constraints, enabling us to determine them without 
explicit knowledge of the bosonic part of Lagrangian. However, 
the necessary calculations are not very illuminating and quite 
tedious already for the dimensionally reduced theory \cite{nic4}.
We also note that (\ref{Susy}) is polynomial in the canonical 
variables, just like the constraints in Ashtekar's formulation of 
$d=4$ gravity.

The canonical brackets given above are sufficient to verify
that the supersymmetry variations (\ref{var248}) and (\ref{varchi}) 
of the basic fields are recovered by means of the formula
\beq
\D_\varepsilon \varphi ({\bf x})= 
\Big\{ \varphi ({\bf x}),  {\mathscr S}[\varepsilon] \Big\}
\eeq
where we have introduced the integrated constraint
\bea
{\mathscr S} [\varepsilon ] :=
\int d {\bf x} \, \oep^I ({\bf x}) {\mathscr S}^{I} ({\bf x})
\eea
with $\varepsilon^I({\bf x})$ an arbitrary spinorial test function.
The verification is equally straightforward for (\ref{varPm})
(with $X_{mn}=0$), but much more tedious for the $\mu=i$ components 
of (\ref{varPmu}) and (\ref{varQmu}), where have performed only 
partial checks. In particular, we are led to postulate non-vanishing
brackets between $P_i^A$, $Q_i^{IJ}$ and $A_i{}^\alpha$ that
have a rather complicated structure due to the occurrence of 
dualized field strengths ``inside'' the connections. However,
these complications are restricted to the components with $\mu=i$ 
and disappear altogether in the reduction to $d=3$.

The $E_8$ structure exhibited by the bosonic sector remains an 
ill understood feature. Of course, $E_8$ is not a symmetry of 
the theory in eleven dimensions, but it does become a (rigid) symmetry 
upon dimensional reduction to three dimensions \cite{julia, marcus}. 
Although the relation of $E_8$ with the internal $d=8$ coordinate 
transformations, which are no longer manifest in the present 
formulation, remains somewhat mysterious, we can offer the following
hints. Returning to the constraint (\ref{dual1}) (its $\mu =t$ 
component, to be precise) and its hamiltonian analog, we recall 
that such constraints need only hold weakly, i.e. on the 
constraint surface. Indeed, checking the compatibility 
of the brackets (\ref{bos3}) and (\ref{bos5}) with the
duality constraint (\ref{dual1}), a little calculation 
reveals that this constraint is modified by a term proportional 
to the SO(16) generator $\phi^{IJ}$: 
\bea
 e^m_A \PtA  - \bruch{1}{2} e^m_{IJ} \phi^{IJ} = 
4  \varepsilon^{ij}\DC_i B_j{}^m \label{cancon}
\eea
Now we recall from \cite{nic4} the identification of the 248-bein 
with the $\sigma$-model field $\VC\in E_8$ appearing in 
the reduction to $d=3$, 
\bea 
e^m_{IJ} = \bruch{1}{60}{\rm Tr} \, \big( Z^m \VC X^{IJ} \VC^{-1} \big)
\quad , \quad
e^m_A = \bruch{1}{60}{\rm Tr} \, \big( Z^m \VC Y^A \VC^{-1} \big)
\eea
where $X^{IJ}$ and $Y^A$ are the compact and non-compact 
generators of $E_8$, respectively, and where the $Z^m$ for 
$m=3,...,10$ are eight nilpotent, hence non-compact, commuting 
generators\footnote{In \cite{nic4} it was incorrectly claimed
that the generators $Z^m$ belong to the Cartan subalgebra of $E_8$. 
This is ruled out by the property ${\rm Tr} (Z^m Z^n) =0$ for all 
$m$ and $n$, which is required for (\ref{id1}) to be satisfied. 
The existence of (at least) eight generators with 
these properties follows e.g. from the decomposition 
${\bf 248}={\bf 64} \oplus {\bf 56} \oplus \overline{\bf 56} \oplus {\bf 28} 
\oplus \overline{\bf 28} \oplus {\bf 8} \oplus \overline{\bf 8}$
of $E_8$ w.r.t. its U(8) subgroup \cite{Murat}.}. On the other hand, 
the $E_8$ Noether charge density of the dimensionally reduced 
theory is given by \cite{nic4,MN}
\bea
\JC = \VC \Big( \PtA Y^A  -\bruch{1}{2} \phi^{IJ} X^{IJ} \Big) \VC^{-1}
\eea 
But in view of (\ref{cancon}), this implies that
\bea
\varepsilon^{ij} \DC_i B_j{}^m 
      = \bruch{1}{60}{\rm Tr} \big( Z^m \JC \big)     ;
\eea
Consequently, the projection of the charge density onto the
nilpotent subalgebra spanned by the $Z^m$'s survives the 
decompactification and is given by $\varepsilon^{ij} \DC_i B_j{}^m$. 
While the eight internal coordinates are thus associated
with the nilpotent subalgebra of $E_8$, the remaining 
part of the $E_8$ Lie algebra presumably corresponds to
some kind of non-commutative geometry.

It does not appear that the present version of $d=11$ supergravity
can be quantized any more easily than the original one of \cite{CJS}. 
However, it is anyhow very unlikely that this theory can be consistently
quantized all by itself: rather, to achieve consistency it must be 
embedded in some as yet unknown, but bigger theory (M-theory?). 
We would thus expect that the remaining open problems can only be 
resolved in such a larger framework. While the ``internal'' sector 
of $\so$ invariant $d=11$ supergravity exhibits a certain conceptual 
simplicity, complications persist in the $d=3$ sector, as can be 
immediately seen by comparing (\ref{varPmu}),(\ref{varQmu}) with
(\ref{varPm}). Since the same might have been said 
about the $d=7$ and $d=4$ sectors of the $\su$ invariant version of 
$d=11$ supergravity given in \cite{dewnic1}, the natural next step
is to search for yet another version of $d=11$ supergravity 
based on a $2+9$ split, which we would expect to possess 
local SO(1,1)$\, \times \,{\rm SO}(16)^\infty$ invariance, where 
${\rm SO}(16)^\infty$ is the maximal compact subgroup of $E_9$ 
defined by means of the generalized Cartan Killing metric 
on the affine Lie algebra $E_9$. The embedding into a bigger 
theory with even larger symmetries might not only explain the 
emergence of space time symmetries from a pre-geometrical theory, 
but should also provide a simplifying principle that might help to 
avoid some of the cumbersome calculations of \cite{dewnic1,nic1,Melosch}.

\medskip
\noindent
{\bf Acknowledgments:} S.~Melosch would like to thank Evangelisches
Studienwerk e.V., Villigst for financial support during part of this work.

%
\end{document}